\documentclass[superscriptaddress]{revtex4}
\usepackage[german,swedish,english]{babel}
\usepackage[dvips]{graphicx}
\usepackage{amssymb}
\usepackage{amsmath}
\usepackage{color}
\newcommand{\bm}{\mathbf}

\begin{document}

\title{The magnetic multi-$\bm{k}$ form factor: an experimental signature of quantum correlators in USb$_{0.88}$Te$_{0.12}$?}

\author{E. Blackburn}
\affiliation{European Commission, JRC, Institute for
   Transuranium Elements, Postfach 2340, Karlsruhe, D-76125 Germany\\}
\affiliation{Institut Laue-Langevin,
   Bo\^\i te Postale 156, F-38042 Grenoble, France\\}
\author{N. Bernhoeft}
\email{nick_bernhoeft@yahoo.com}
\altaffiliation{Present Address: 18 Maynestone Road,
Chinley, SK23 6AQ, United Kingdom.}
\affiliation{DRFMC, CEA-Grenoble, F-38054 Grenoble, France\\}
\date{\today}

\begin{abstract}
Magnetic diffraction patterns which exhibit more than one discrete set of symmetry related peaks have been explained by the simultaneous coexistence of more than one magnetic polarisation and propagation vector, $\bm{k}$, in the scattering amplitude. In this work we suggest using such multi-$\bm{k}$ magnetic structures as prototypical multiple order parameter systems.  The examples studied are uranium rocksalts, which exhibit magnetic diffraction patterns, the indexing of which requires three orthogonal propagation wavevectors.  In the cubic phase these systems exhibit, in addition to the three sets of symmetry related magnetic diffraction peaks, a set of additional Bragg peaks which have been observed by both neutron and x-ray resonant scattering.  An interpretation of the neutron diffraction form factor of these peaks is presented, identifying these peaks as signatures of quantum coherence between the three orthogonal order parameters.
\end{abstract}

%\pacs{}

\maketitle

\section{Introduction}

Experimental probes of quantum correlations are of fundamental interest in developing an understanding of the many-body state. Active areas of research include few-atom Bose condensates and strongly correlated electron systems exhibiting, for example, joint superconducting-magnetic phases. This paper, a sequel to a series of experimental papers \cite{LPBL02,BPDW+04,LB04,BBMW+06}, addresses the microscopic expression of quantum correlation in bulk condensed matter through a direct calculation of the neutron scattering amplitude for both single-$\bm{k}$ ($\bm{k}_1$) and triple-$\bm{k}$ ($\bm{k}_3$) diffraction from the non-coplanar magnetic configurations of the cubic rock salt compounds USb$_{0.82}$Te$_{0.18}$ and UAs$_{0.8}$Se$_{0.2}$.

Whilst the $\bm{k}_1$ intensities can be interpreted within the conventional cross section, for the $\bm{k}_3$ terms there are two distinct problems. Firstly, given that the triple-$\bm{k}$ scattering amplitude lies outside the conventional cross section, the existence of the long-range-ordered signal needs an explanation; and secondly, the highly structured neutron diffraction form factor requires interpretation. The first point has been addressed in earlier work \cite{BB05}, where it is shown that the enigmatic triple-$\bm{k}$ peaks may be understood as two-site auto-correlations of a quantum correlator. This paper focusses on inferences that may be drawn from numerical simulations of the relative Bragg neutron diffraction intensities.

The wealth of information contained in the highly anisotropic $\bm{k}_3$ form factor is interpreted within a generalised electromagnetic scattering amplitude founded in Clifford geometric algebra \cite{BB05,DL03}. The inferred electronic density distribution is an important clue, and constraint, on the physical origin of such $\bm{k}_3$ peaks \cite{note}.

Assuming that the magnetic interaction is dominated by orbitals of uranium 5$f$ symmetry consistent with a cubic environment, it is found that an interpretation of both the single- and triple-$\bm{k}$ form factors is (uniquely) possible via the degenerate set of 5$f$($\delta$) orbitals of the form $5f(\delta) = \{(5x^3 - 3xr^2)/r^3, (5y^3 - 3yr^2)/r^3, (5z^3 - 3zr^2)/r^3\}$.

\section{Experimental motivation}

The motivation for this study has arisen from the experimental investigation of non-coplanar, i.e. multi-$\bm{k}$, magnetic configurations. These have been examined by x-ray resonant scattering and neutron diffraction techniques in the uranium rock-salts UAs$_{0.8}$Se$_{0.2}$, USb$_{0.88}$Te$_{0.12}$ and USb$_{0.85}$Te$_{0.15}$ \cite{LPBL02,BPDW+04,LB04,BBMW+06} as prototype multiple order parameter systems with a view towards understanding, for example, possible joint magnetic-superconducting phases and novel spin, charge and dynamic states invoking high-order quantum correlations. 

Multi-$\bm{k}$ configurations may be understood as an extension of the familiar single-$\bm{k}$ ($\bm{k}_1$) states found in ferromagnets and other coplanar magnetic structures (e.g. N\'eel antiferromagnets and helical magnetic structures) which are described by single particle projectors of the state vector yielding magnetic order parameters of the form, $\mathcal{M}_{\bm{k}_{\alpha}} = M_{\bm{k}_{\alpha}} \exp(i \theta_{\bm{k}_{\alpha}})$, where $\alpha$ is a directional coordinate, $M_{\bm{k}_{\alpha}}$ is the polarisation vector and $\theta = \bm{k}_{\alpha} \cdot \bm{r}_i + \theta_{\alpha}$ is the relative phase at the $i^{\mathrm{th}}$ position. Such $\bm{k}_1$ magnetic configurations are specified by one polarisation, $M_{\bm{k}_{\alpha}}$, and one propagation wavevector, $\bm{k}_{\alpha}$.

More intricate magnetic structures may be described by a set of state vector projectors, invoking more than one order parameter at a given site (a multi-$\bm{k}$ configuration). For example, a 3-$\bm{k}$ configuration, which yields 3 independent sets of magnetic Bragg peaks around a given reciprocal lattice vector $\tau$ corresponding to three (orthogonal) propagation wavevectors, requires a set of orthogonal terms, $\{\mathcal{M}_{\bm{k}_{\alpha}} = M_{\bm{k}_{\alpha}} \exp(i \theta_{\bm{k}_{\alpha}}), \mathcal{M}_{\bm{k}_{\beta}} = M_{\bm{k}_{\beta}} \exp(i \theta_{\bm{k}_{\beta}}), \mathcal{M}_{\bm{k}_{\gamma}} = M_{\bm{k}_{\gamma}} \exp(i \theta_{\bm{k}_{\gamma}})\}$ with the polarisation and propagation vectors defining three independent order parameters \cite{LPBL02,BPDW+04,LB04,BBMW+06}. 

The rocksalts studied exhibit an antiferromagnetic 3-$\bm{k}$ configuration. The positions and relative intensities of the $\bm{k}_1$ Bragg peaks are described by a longitudinal (++-- --) moment configuration along each of the principal cubic axes, with ordering wave vectors $\bm{k}$ = $\langle$0 0 1/2$\rangle$ reciprocal lattice units (rlu). The puzzle in these materials is the consistent appearance, in both neutron and x-ray resonant diffraction, of a fourth set of Bragg peaks corresponding with a T$_{1u}$ symmetry element (c.f.~magnetic dipole) with propagation wavevector $\bm{k}_3$ where $\bm{k}_3$ = $\bm{k}_x$ + $\bm{k}_y$ + $\bm{k}_z$. Although the $\bm{k}_3$ peaks are typically between 3 and 4 orders of magnitude weaker than those arising from the primary order parameters at $\bm{k}_1$ positions, the body of experimental data suggests that they are not artefacts or multiple scattering resonances \cite{LPBL02,BPDW+04,LB04,BBMW+06}. Their long range order, temperature, azimuth and electric dipole [E$_1$-T$_{1u}$] x-ray photon energy dependencies attest to them representing a 'magnetic' dipole term with relative phasing (+--+--) and ordering wavevector $\bm{k}_3$ = $\langle$1/2 1/2 1/2$\rangle$ rlu.

\section{Analysis of the $\bm{k}_1$ and $\bm{k}_3$ neutron diffraction form factors}

The $\bm{k}_3$ diffraction resonances, requiring simultaneously the vector sum and direct product of the three underlying magnetic order parameters to generate the effective dipole and phase respectively, lie outside the conventional two-site (level-1) correlation function as sampled by both neutron and [E$_1$-T$_{1u}$] x-ray diffraction techniques \cite{BB05,DL03,LB89}. 

Inspection of the neutron data reveals the systematic absence of neutron diffraction peaks both at $\bm{Q}_{\bm{k}_1} = \langle$0 0 $n$/2$\rangle$ and $\bm{Q}_{\bm{k}_3} = \langle n$/2 $n$/2 $n$/2$\rangle$ for $n$ = odd integer. The first observation leads to a magnetic structure having a $\langle$001$\rangle$ dipole moment parallel to $\bm{k}_1$ with (++-- --)phasing \cite{LPBL02,BPDW+04,LB04,BBMW+06}; the second suggests an alternating (+--+--) longitudinal $\langle$111$\rangle$ moment along $\bm{k}_3$ \cite{BPDW+04,BBMW+06}.  The two demands are incompatible and a central result of this paper is that such a picture is indeed untenable in light of the anisotropy and wavevector dependence of the $\bm{k}_1$ and $\bm{k}_3$ neutron diffraction intensities.

Before it is possible to make an analysis of the intensity distribution (form factor), the origin of the scattering amplitude must be understood. For the $\bm{k}_1$ peaks there is no problem since they are well explained by the conventional neutron diffraction amplitude. However, the $\bm{k}_3$ peaks lack a clear interpretation \cite{LPBL02,BPDW+04,LB04,BBMW+06,BB05,DL03}. To this end we present a generalised formalism of the scattering amplitude founded in Clifford geometric algebra \cite{BB05,DL03}. 

Fundamental to the approach, which represents spinor matrix mechanics using a geometric algebra based in Euclidean 3-D space, is the translation of the set of spinor matrices, $\{\sigma_x, \sigma_y, \sigma_z\}$, to the set of $\{\bm{e}_x, \bm{e}_y, \bm{e}_z\}$ Euclidean axes. On introduction of a geometric product, $e_i e_j = \delta_{ij} + \sqrt{-1} \epsilon_{ijk} e_k$, an algebra of 8 distinct elements, $\{1; e_i,e_j,e_k ; e_ie_j, e_je_k,e_ke_i; I \}$, is generated where the product of two orthogonal vectors is called a bivector ($bv$) and $I$, the unit trivector, is the geometric product of three orthogonal vectors. 

This reformulation of Pauli matrix algebra provides a basis for the systematic analysis of space-time correlations of extended systems \cite{BB05}. A key concept is the projection of the (mesoscopic) multi-particle state-vector $\Psi$ onto a set of subspaces. For single quasiparticle states, $N_1$ independent subspaces are identified with a set of $N_1$ frames of quantisation (level-1 quantum correlation). For joint `two particle' or level-2 correlators there are a set of $N_2$ joint subspaces and so forth. 

The spinor, $\Psi(\bm{x},t)$, acts as a time dependent field over $\bm{x}$ serving to rotate the external reference frame to the local orientation of the frame of quantisation. A distinct frame of quantisation, i.e.~set $\{\bm{e}_{\bm{k}}^p\}$, is associated with each (quasi)particle state. An incoherent set of $N_1$ states is then represented by $N_1$ frames of quantisation of arbitrary relative orientation, $\beta^{p,q}$, between the $\bm{e}_3^p : \bm{e}_3^q$ axes. This is a representation of the disordered state. The transition to long-range order may then be represented by one preferred frame of quantisation where the condensation of $\beta^{p,q}$ breaks the rotational symmetry of the paramagnet state. The quantisation axis of this global frame, $\{\bm{e}_{\bm{k}}^{OP1}\}$, abstracts the essential geometry of level-1 long range order and may be taken as the microscopic order parameter enabling great conceptual and computational simplification.  

Extension to states of multiple electronic order invokes an additional phasing variable, $\gamma_{\bm{k}}^{i,j}$, as a measure of the phase between elements of the order parameter sets, $\{\bm{e}_{\bm{k}}^{OPi}\} : \{\bm{e}_{\bm{k}}^{OPj}\}$. The $\bm{k}_3$ relative intensities provide evidence that the fourth set of Bragg peaks arises through the $\gamma_{\bm{k}}^{i,j}$ phase coherence of the underlying set of single-k order parameters \cite{LB04,BB05}. 

\section{Neutron scattering}

Within Clifford algebra physical observables are expressed as the geometric product of an appropriate state vector with a hierarchal set of level-$n$ quantum correlators, $\{E_n, J_n\}$ \cite{DL03}. For a level-1, i.e.~single quasiparticle representation of $\Psi$, one has $E_1 \sim 1$ and $J_1 \sim \mathcal{M}^z_{\bm{k}_{\alpha}}$ where the $\sim$ sign indicates 'behaves as'. The magnetic cross-section is then separated into an interaction together with the quantum-statistical expectation of the (single-particle) magnetic auto-correlation $\sim \langle J_1(r,t) J_1(r',t') \rangle$. Of particular interest to the present work is the level-3 auto-correlation function $\sim \langle J_3(r,t) J_3(r',t') \rangle$ where $J_3 \sim (\mathcal{M}^z_{\bm{k}_{\alpha}} + \mathcal{M}^z_{\bm{k}_{\beta}} + \mathcal{M}^z_{\bm{k}_{\gamma}} - \mathcal{M}^z_{\bm{k}_{\alpha}} \mathcal{M}^z_{\bm{k}_{\beta}} \mathcal{M}^z_{\bm{k}_{\gamma}})$ \cite{BB05}.

The neutron-electron dipole-dipole scattering amplitude arises from an effective Zeeman field. The magnitude depends on the relative separation and spin configuration of the two particles. Written for an electron sitting in the neutron field,
\begin{equation}
\hat{H}_i = (g \mu_B \gamma \mu_N ) \hat{\sigma}_i \cdot \nabla \times \Big( \frac{\hat{s}\times R}{\vert R \vert ^3} \Big)
\end{equation}
where $\hat{\sigma}$ and $\hat{s}$ are the electron and neutron spin operator respectively and the remaining symbols take on their conventional meanings \cite{LB89}. Geometrical restrictions, imposed by the form of the electromagnetic interaction, generate powerful selection rules. To exploit these it is convenient to express $R/\vert R \vert ^3$ as $\nabla (1/\vert R \vert )$ and subsequently decompose $(1/\vert R \vert )$ in plane wave states (we keep the conventional symbol $i = \sqrt{-1}$ in the expression of harmonic functions) giving,
\begin{equation}
\hat{H}_i = 4 \pi \hat{\sigma}_i \cdot \lbrack \kappa \times (\hat{s} \times \kappa) \rbrack \frac{e^{i\kappa \cdot r_i}}{\kappa^2}.
\end{equation}
This expression is microscopic and symmetric both with respect to wavevector and $\hat{\sigma}$ and $\hat{s}$ of the electron and neutron respectively and may be interpreted as the Hamiltonian of an electron in the effective Zeeman field of the neutron. 

Mapping this two level system to GA gives the electron-neutron dipole interaction term,
\begin{equation}
\mu_B B_k \langle \psi \vert \hat{\sigma}_k^{OPj} \vert \psi \rangle \leftrightarrow \mu_B B_k \lbrack \langle \psi^{\dagger} I e_k^{OPj} \psi I e_3^{OPj} \rangle - \langle \psi^{\dagger} I e_k^{OPj} \psi \rangle I e_3^{OPj} \rbrack
\end{equation}
where the left-hand side is the conventional Pauli-matrix notation and the right-hand side the GA equivalent. Since the second element of the right-hand bracket has no scalar part, 
\begin{eqnarray}
\mu_B B_k \langle \psi \vert \hat{\sigma}_k^{OPj} \vert \psi \rangle & \leftrightarrow & \mu_B B_k \langle I e_k^{OPj} \psi I e_3^{OPj} \psi^{\dagger} \rangle \nonumber \\
& \leftrightarrow & (\mu_B I B_k ) \cdot (\psi J_1 \psi^{\dagger})_{bv} 
\end{eqnarray}
where $(\psi J_1 \psi^{\dagger})_{bv}$ is the bivector projection of the enclosed geometric product \cite{BB05,DL03}. The interaction may be thus viewed as the bivector projection of the (electron) multivector $\psi J_1 \psi^{\dagger}$ on the (neutron's) magnetic field. This formalism forms a basis for interpreting the scattering amplitude from many electron states at the level-1 projection of the state vector. Explicitly, $J_1 \equiv I e_3^{OPj}$ is the axial vector dual of the $e_1 e_2$ bivector of the $j^{\mathrm{th}}$ order parameter $OP_j$ and the measured autocorrelation function is $\langle \lbrack I \bm{e}_z^{OP1} \cdot (\psi J_1 \psi^{\dagger}) \rbrack_{(r,t)} \lbrack I \bm{e}_z^{OP1} \cdot (\psi J_1 \psi^{\dagger}) \rbrack_{(r',t')} \rangle$which is read as the $(r,t) \leftrightarrow (r',t')$ correlation of the bivector projection of the level-1 quantum correlator, i.e. $(\psi J_1 \psi^{\dagger})$, on the quantisation axis of the order parameter subspace labelled 1.

The dipole cross-section of the two-site $J_1$ correlator permits inferences on collinear and single order parameter coplanar configurations to be made. However, higher order coherencies within the state vector, such as those present in 3-$\bm{k}$ systems, are not represented, and the conventional formalism yields no clear way forward. 

The strength of the present formalism is then the immediate generalisation to the level-3 correlator, to give the interaction basic to the scattering amplitude, $(\mu_B I B_k ) \cdot (\psi J_3 \psi^{\dagger})_{bv}$. We suggest to calculate the scattering amplitude for the diffraction peaks in the 3-$\bm{k}$ state through this $J_3$ projection of the state vector and offer, as empirical validation, the success of this formalism in accounting for the relative intensities and intricate anisotropies of the $\bm{k}_3$ peaks (see Figure 2 and discussion below). 

In the presence of periodic translational symmetry the $J_3$ correlator is taken to have the following form \cite{DL03}, 
\begin{equation}
J_3 = \frac{1}{4} (I \bm{e}_3^{OP1} e^{ik_x.r} + I \bm{e}_3^{OP2} e^{ik_y.r} + I \bm{e}_3^{OP3} e^{ik_z.r} - I \bm{e}_3^{OP1} e^{ik_x.r} I \bm{e}_3^{OP2} e^{ik_y.r} I \bm{e}_3^{OP3} e^{ik_z.r})
\end{equation}
where the relative inter-order-parameter phase has been set to zero, the three characteristic wave vectors of long range order are $\bm{k}_x$, $\bm{k}_y$ and $\bm{k}_z$ respectively, and $I\bm{e}_3^{OPj}$ is the abbreviation of $I^{OPj}\bm{e}_3^{OPj}$. 

As argued above, the symmetries of the $J_1$ peaks in the 3-$\bm{k}$ phases of the uranium rock-salts force the representative moments in the scattering amplitude to be longitudinal, i.e.~parallel to the appropriate $\bm{k}_1$ vector. The $j^{\mathrm{th}}$ subspace axis of quantisation, $\bm{e}_3^{OPj}$, is therefore taken parallel to its $\bm{k}_1$ vector in keeping with the cubic symmetry, and the last term of $J_3$ then automatically gives rise to the stationary (Bragg) phase condition for the fourth set of Bragg peaks at $\bm{k}_3 = \bm{k}_x+\bm{k}_y+\bm{k}_z$. 

\section{Calculation of $\bm{k}_1$ and $\bm{k}_3$ Bragg peaks}

With this as a basis, the relative intensities and anisotropies of the scattering amplitudes for both $\bm{k}_1$ and $\bm{k}_3$ Bragg events have been calculated, forming the relevant matrix elements from the 5$f$ wavefunctions, which split in a cubic environment into a singlet and two triplets labelled $\beta$, $\delta$ and $\epsilon$ respectively. In the calculations the point group symmetries of the ionic orbitals in a cubic environment have been respected; however, the radial extent of the orbitals differs from the free ion values in the metallic state. To assess the spatial extent of the 5$f$ wavefunctions a calculation of the $\bm{k}_1$ Bragg intensities has been made in which the set $\{j_0, j_4, j_6\}$ of ionic Bessel functions compatible with cubic symmetry \cite{note2} are replaced by a single term, $\exp(-0.035 \vert \bm{Q} \vert)$ with $\vert \bm{Q} \vert$ in \AA $^{-1}$. This scheme, which replaces $\langle j_0 \rangle$ by $\exp(-0.035 \vert \bm{Q} \vert)$ whilst setting the $\langle j_4 \rangle$ and $\langle j_6 \rangle$ coefficients to zero, is used to represent the radial extent of the uranium ion in rock-salt compounds \cite{LMSV76,RLB84}.  Since the $\langle j_0 \rangle$ coefficient is independent of the orientation of the scattering vector with respect to the state vector, the approximation precludes any information being obtained on the angular disposition of the wavefunction. 

The results are given as closed circles in the left-hand panel of Figure 1. This single term approximation to the $\{j_0, j_4, j_6\}$ set reproduces the wave vector dependency of the measured intensities, and enables one, in a semi-empirical manner, to address the role of bonding on the radial extent of the uranium valence shell. The lack of numerical agreement for the first two data points is attributed to extinction in this large (700 mg) crystal. The figure further demonstrates that the marked anisotropies at $\vert \bm{Q} \vert \sim 2.9$ \AA $^{-1}$, 3.8 \AA $^{-1}$ and 4.1 \AA $^{-1}$ are correctly reproduced by the complex vector scattering amplitude of the level-1 projection of the state vector. The right-hand panel of Figure 1 is the transform of the integrated intensities of the left-hand panel, i.e.~both experimental (open circles) and theoretical (closed circles), to a form factor together with the extinction corrected (open triangles) form factor of pure USb \cite{LMSV76}. The latter has been used to set the integrated intensity on an absolute scale.

Moving now to the $\bm{k}_3$ peaks, we note that similar, anomalous, $\vert \bm{Q} \vert$ dependent intensities of the $\bm{k}_3$ reflections are observed in UAs$_{0.8}$Se$_{0.2}$ \cite{LPBL02,BPDW+04,LB04,BBMW+06} and the conclusions drawn are hence likely to be generic to this class of 3-$\bm{k}$ state. From the relative intensities of $\bm{k}_1$ Bragg peaks in UAs$_{0.8}$Se$_{0.2}$ and USb$_{0.88}$Te$_{0.12}$, taken at $\vert \bm{Q} \vert$ above 2.5 \AA $^{-1}$, where extinction corrections appear minimal (see Figure 1), a scaling of the UAs$_{0.8}$Se$_{0.2}$ data (open triangles) to the USb$_{0.88}$Te$_{0.12}$ data (open circles) has been made. Using this $\bm{k}_1$ scaling, the $\bm{k}_3$ points of both compounds have been amalgamated in Figure 2. 

Direct calculation shows the minimal $\lbrack \langle j_0 \rangle \rightarrow \exp(-0.035 \vert \bm{Q} \vert); j_4 = j_6 = 0 \rbrack$ approximation is insufficient to explain the intensity-wavevector distribution of the $\bm{k}_3$ peaks. Therefore the summed contribution from the radial $\{j_0, j_4, j_6\}$ set as determined by the cubic environment is employed at each angular setting. As with the $\bm{k}_1$ peaks, ignoring the role of valence shell bonding results in $\bm{k}_3$ intensities that have poor agreement with the data since the radial extent of the wavefunction is artificially restricted. The same $\bm{k}_1$ replacement of the $\langle j_0 \rangle$ term is therefore made to allow for the bond electron density redistribution in the radial extent of the isotropic component, and, in the same vein, the calculations make use of an, empirical, adjustment of the relative $\langle j_4 \rangle$ contribution. 

Calculations of the scattering amplitude arising from the $\{ \delta \}$, $\{ \epsilon \}$ and $\{ \beta \}$ sets of symmetry determined orbitals in a cubic environment give a ratio of cross sections: 5$f$($\delta$):5$f$($\epsilon$):5$f$($\beta$)::30000:100:1 in favour of 5$f$($\delta$). Furthermore, the coherent scattering amplitude from the 5$f$($\delta$) orbitals is uniquely able to reproduce the observed anisotropy of the cross section. Most notably this is the case between the $\{$(5/2 5/2 1/2), (7/2 1/2 1/2)$\}$ and $\{$(5/2 5/2 3/2), (7/2 3/2 1/2)$\}$ pairs of reflections at $\vert \bm{Q} \vert$ = 3.526 \AA $^{-1}$ and $\vert \bm{Q} \vert$ = 3.900 \AA $^{-1}$ respectively. The anomalies are captured by the 5$f$($\delta$) orbitals whilst the 5$f$($\beta$), 5$f$($\epsilon$) or a spherically symmetric linear combination of the orbitals all give the $\vert \bm{Q} \vert$ = 3.526 \AA $^{-1}$ anisotropy in the opposing sense. Thus, despite systematic absences at $\bm{Q} = \langle n$/2 $n$/2 $n$/2 $\rangle$ for $n$ odd integer and the $\{ xyz / r^3 \}$ symmetry, the $\beta$-singlet appears to play no significant role in the 3-$\bm{k}$ neutron diffraction and conceptual images of a 'moment' along $\langle$111$\rangle$ directions with $x$,$y$,$z$ 'projections' may be misleading.  

A comparison of the relative cross sections for $\bm{k}_1$ and $\bm{k}_3$ peaks held within the $J_3$ projector for the 5$f$($\delta$) orbitals have a ratio, $\bm{k}_3$:$\bm{k}_1$ :: 50:1. From the empirical intensity ratio of Bragg peaks, $\bm{k}_3$:$\bm{k}_1$ :: 1:10$^3$ (see Figures 1 and 2), one infers the $J_1$ contribution to the $\bm{k}_1$ peaks is dominant. This observation underlies the successful use of the standard cross-section in previous interpretations. The calculated ratios are collated in Table 1.

\begin{table}[htb]
\begin{center}
\begin{tabular}{|c|c|}
\hline
$J_3$ correlator & Relative intensity at $\bm{Q} = \tau + \bm{k}_n$ \\
\hline
5$f$($\delta$) triplet ($\bm{k}_3$) & 1 \\
5$f$($\epsilon$) triplet ($\bm{k}_3$) & $3 \cdot 10^{-3}$ \\
5$f$($\beta$) singlet ($\bm{k}_3$) & $3 \cdot 10^{-5}$ \\
5$f$($\delta$) triplet ($\bm{k}_1$) & $2 \cdot 10^{-2}$ \\
\hline
\end{tabular}
\caption{Comparative Bragg peak intensities of the $J_3$ correlator.  }
\end{center}
\label{tab:Table1}
\end{table}

\begin{figure}[tbhp]
\begin{center}
\includegraphics[width=12.0cm]{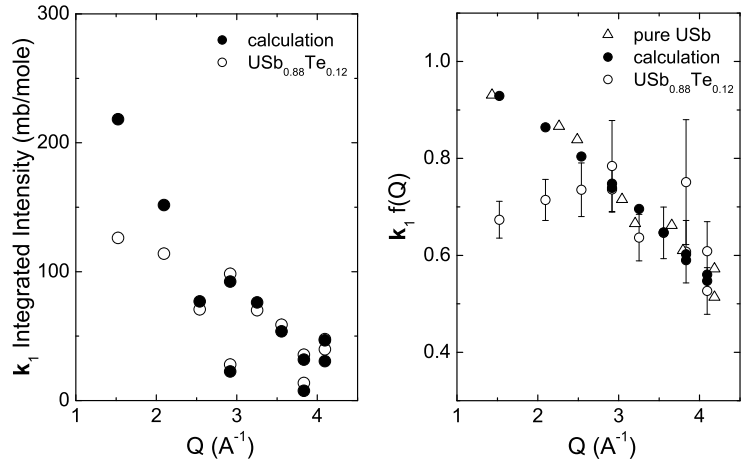}
\end{center}
\caption{Left-hand panel: $\bm{k}_1$ scattering intensity compared with a first principles calculation (excepting a vertical scaling factor) from the 5$f$($\delta$) orbitals compatible with cubic symmetry. Right-hand panel: data converted to form factor, additional triangular points are taken from the observed form factor of pure USb \cite{LMSV76}. The latter have been used to set the integrated intensity on an absolute scale in the left hand panel (mb/mole). Key: open circles = USb$_{0.88}$Te$_{0.12}$; open triangles = USb; closed circles = calculation. The USb$_{0.88}$Te$_{0.12}$ data are not corrected for extinction. }
\label{fig:Fig1}
\end{figure}

\begin{figure}[bhp]
\begin{center}
\includegraphics[width=6.0cm]{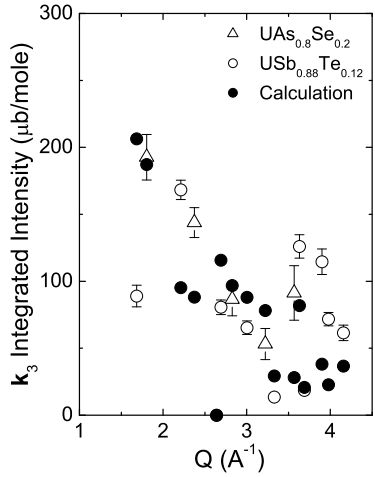}
\end{center}
\caption{$\bm{k}_3$ scattering intensity compared with a first principles calculation (excepting a vertical scaling factor) from the degenerate 5$f$($\delta$) orbitals compatible with cubic symmetry. Key: open circles = USb$_{0.88}$Te$_{0.12}$; open triangles = UAs$_{0.8}$Se$_{0.2}$; closed circles = calculation. The $\{$(5/2 5/2 1/2), (7/2 1/2 1/2)$\}$ and $\{$(5/2 5/2 3/2), (7/2 3/2 1/2)$\}$ pairs of reflections at $\vert \bm{Q} \vert$ = 3.526 and $\vert \bm{Q} \vert$ = 3.900 \AA $^{-1}$ respectively have been separated by 2\% in their abscissae to illustrate the anisotropy. The data are set on an absolute scale as in Figure 1.}
\label{fig:Fig2}
\end{figure}

\section{Implications and conclusion}

The Clifford level-3 quantum correlators, interpreted as projectors of the mesoscopic state vector, yield a scattering amplitude comprising both a sum and geometric product of three independent order parameters \cite{BB05}, permit a consistent interpretation of the observed peaks in terms of three coherent scattering amplitudes coming from the 5$f$($\delta$) triplet state. The magnetic-dipole terms of $(\psi J_n \psi^{\dagger})_{bv}; n = 1,3$ are then extracted by the projection effected by the scattering amplitude onto 3 orthogonal subspaces which may be identified by their respective order parameters. The individual and summed scattering amplitudes over the 5$f$($\delta$) subspaces then determine the $\bm{k}_1$ and $\bm{k}_3$ cross sections respectively. The present calculations of the neutron dipole form factor show that the origin of the 3-$\bm{k}$ Bragg peaks is consistent with selection of the $I \mathrm{e}_z^{(1)} I \mathrm{e}_z^{(2)} I \mathrm{e}_z^{(3)}$ component of the $J_3$ quantum correlator through the \emph{wavevector selectivity} of coherent diffraction, i.e.~fulfilling the Bragg condition at $\bm{k}_3$ = $\langle kkk \rangle$. Further, the overwhelming relative amplitude of the 5$f$($\delta$) triplet and its unique ability to explain the observed anisotropy indicate that it is the most plausible candidate as the orbital basis to understand the $\bm{k}_3$ diffraction peaks within this formalism.

The existence and anisotropy of $\bm{k}_3$ Bragg reflections give unique evidence for the space-time long range order of a level-3 projection of the state vector. The advantage of the magnetic system studied here is the direct observation afforded by neutron diffraction. Associating a set of 5$f$($\delta$) orbitals with each uranium site, the present work suggests that: (i) valence shell bonding gives a radial density redistribution within the cubic point group symmetry, and (ii) the electronic system sits in a spatially correlated coherent superposition of the set of $\{ \pm \delta \lbrack x \rbrack, \pm \delta \lbrack y \rbrack, \pm \delta \lbrack z \rbrack \}$ orbitals. In particular, the $\bm{k}_3$ null result at $\langle n$/2 $n$/2 $n$/2$\rangle$ for $n$ odd integer leads to the conclusion that the set $\{ \mathcal{M}_{\bm{k}_{\alpha}}^z, \mathcal{M}_{\bm{k}_{\beta}}^z, \mathcal{M}_{\bm{k}_{\gamma}}^z \}$ of vector projected order parameters are aligned with the $\{ \pm Q_x, \pm Q_y, \pm Q_z \}$ axes respectively. Thus, the relative phase of the order parameter set is 'ferromagnetic' and it follows that $\{ \mathcal{M}_{\bm{k}_{\alpha}}^z, \mathcal{M}_{\bm{k}_{\beta}}^z, \mathcal{M}_{\bm{k}_{\gamma}}^z \}$ and $\{ \pm Q_x, \pm Q_y, \pm Q_z \}$ share the same chirality. 

There is a distinction between the scattering state vector, $\Psi$, and the scattering amplitude as the bivector projection of $(\Psi J_n \Psi^{\dagger})$. For $\bm{k}_1$ peaks, whether they derive from $J_1$ or $J_3$ quantum correlations, the difference is not apparent. It comes to light for $\bm{k}_3$ peaks which derive from spatially extended correlations of the product term associated with the level-3 quantum correlator. The $J_3$ correlator projects three scattering amplitudes associated with the three independent Pauli subspaces. 

These three amplitudes give rise to two distinct sets of diffraction peaks associated with the two level-1 and level-3 projections of the state vector. It is suggested that the two sets may be identified with two types of electronic order, (antiferro)magnetic (++-- --) $\bm{k}_1$ and (antiferro)chiral (+--+--) $\bm{k}_3$ order respectively. The notion of $\bm{k}_1$ as 'primary' and $\bm{k}_3$ 'secondary' order parameters, based on the respective thermal evolution of the associated sets of Bragg reflections, may in this respect be a result of the temperature dependencies of the $J_1$ and $J_3$ correlators respectively.

It would appear difficult for arguments based on semi-classical approximations to level-1 projections of $\Psi$ to capture the essence of the triple-$\bm{k}$ quantum state, and by implication, of the non-coplanar magnetic state in general. The relevant correlations appear to lie outside of such a formalism. A similar limitation may confound intrinsically level-1 approaches to problems such as the joint superconductor-magnetic phases and quantum critical point scenarios, which are of great topical interest. 

The order parameter concept is general. A particular advantage of the present realisation of multiple order parameters as magnetic states, where $\Psi$ is encoded as a rotor giving the space-time local orientation of a spin-vector, is the clear geometric interpretation. We hope that this work will stimulate further research in multiple order parameter systems. The role of the quantum correlators of the mesoscopic state vector may be of direct interest in the study of more complex multiple order parameter systems such as magnetic superconductors and mutual charge, orbital and magnetic ordered states. In particular the present study motivates the search for experimental evidence of both further states of long range order in level-$n$ quantum correlators and the existence of novel states that may be revealed through $\langle J_1 J_3 \rangle$ cross-correlators.


\begin{thebibliography}{1}
\bibitem{LPBL02} M. J. Longfield, J. A. Paixao, N. Bernhoeft and G. H. Lander, {\it Physical Review B} {\bf 66}, 054417 (2002). 
\bibitem{BPDW+04} N. Bernhoeft, J. A. Paixao, C. Detlefs, S. B. Wilkins, P. Javorsky, E. Blackburn and G. H. Lander, {\it Physical Review B} {\bf 69}, 174415 (2004).
\bibitem{LB04} G. H. Lander and N. Bernhoeft, {\it Physica B} {\bf 34}, 345 (2004).
\bibitem{BBMW+06} E. Blackburn, N. Bernhoeft, G. J. McIntyre, S. B. Wilkins, P. Boulet, J. Ollivier, A. Podlesnyak, P. Javorsky, G. H. Lander, K. Mattenberger and O. Vogt, {\it Philosophical Magazine} {\bf 86}, 2553 (2006).
\bibitem{BB05} E. Blackburn and N. Bernhoeft, accepted in {\it J. Phys. Soc. Jpn.} (2006); E. Blackburn, Ph. D. Thesis, Universit\'e Joseph-Fourier, Grenoble I, France (2005).
\bibitem{DL03} C. J. Doran and A. Lasenby, {\it Geometric Algebra for Physicists}, Cambridge University Press, Cambridge (2003); D. Hestenes {\it Space-Time Algebra}, Gordon and Breach, New York (1966).
\bibitem{note} The x-ray resonant scattering amplitude contains no information on the spatial distribution of the diffracting object, and knowledge is limited to what may be obtained from the spatial symmetries of the correlation function as seen through core- to valence-band absorption-emission resonance spectroscopy.
\bibitem{LB89} S. W. Lovesey and E. Balcar, {\it Theory of Magnetic Neutron and Photon Scattering}, Oxford University Press, Oxford (1989).
\bibitem{note2} The $j_2$ term is not allowed in cubic symmetry.
\bibitem{LMSV76} G. H. Lander, M. H. Mueller, D. M. Sparlin and O. Vogt, {\it Physical Review B} {\bf 14}, 5035 (1976).
\bibitem{RLB84} J. Rossat-Mignod, G. H. Lander and P. Burlet, in Vol. 1, p. 415 of the {\it Handbook of the Physics and Chemistry of the Actinides} edited by A. J. Freemand and G. H. Lander, North-Holland, Amsterdam (1984).
\end{thebibliography}
\end{document}